\documentclass[onecolumn,prd,nofootinbib,showpacs]{revtex4}

\usepackage{textcomp}
\usepackage{mathtools, ulem, graphicx}
\usepackage[export]{adjustbox}
\usepackage{sidecap}
\usepackage{verbatim}

\newcommand{\e}{\begin{equation*}\begin{aligned}}
\newcommand{\ee}{\end{aligned}\end{equation*}}

\newcommand{\en}{\begin{equation}\begin{aligned}}
\newcommand{\een}{\end{aligned} \end{equation}}

\newcommand{\p}{\partial}
\newcommand{\pf}[2]{\frac{\p #1}{\p #2}}
\newcommand{\f}[2]{\frac{#1}{#2}}

\newcommand{\ra}{\rangle}
\newcommand{\la}{\langle}

\newcommand{\da}{\dagger}
\newcommand{\ma}{\mathcal}

\newcommand{\Q}{\left}
\newcommand{\W}{\right}

\newcommand{\pma}{\begin{pmatrix}}
\newcommand{\epma}{\end{pmatrix}}

\newcommand{\na}{\nabla}

\begin{document}

\title{Dilational Symmetry-Breaking in Thermodynamics}
\author{Chris L. Lin}
\author{Carlos R. Ord\'{o}\~{n}ez}
\affiliation{Department of Physics, University of Houston, Houston, TX 77204-5005}

\date{\today}
\email{cllin@uh.edu}
\email{cordonez@central.uh.edu}

\begin{abstract}
Using thermodynamic relations and dimensional analysis we derive a general formula for the thermodynamical trace $2\ma E-DP$ for nonrelativistic systems and $\ma E-DP$ for relativistic systems, where $D$ is the number of spatial dimensions, in terms of the microscopic scales of the system within the grand canonical ensemble. We demonstrate the formula for several cases, including anomalous systems which develop scales through dimensional transmutation. Using this relation, we make explicit the connection between dimensional analysis and the virial theorem. This paper is focused mainly on the non-relativistic aspects of this relation.

\end{abstract}


\maketitle

\section{Introduction}

The quantity $2\ma E-DP$ for non-relativistic systems, or $\ma E-DP$ for relativistic systems, where $\ma E$ is the thermal energy density, $D$ the number of spatial dimensions, and $P$ the pressure, plays an important role in physics. For example, this quantity is the thermal analog of the trace of the improved stress-energy tensor which is a measure of dilational symmetry-breaking and which plays a central role in the renormalization group \cite{brown,car,gat}. \\

In non-relativistic physics, $2\ma E-DP$ can be used as a measure of deviations of real gases from ideal ones. Traditionally, such deviations are measured by giving the two systems the same value for two of their thermodynamic variables, and taking the difference between them for a third. For example, the difference in pressure between a real gas and an ideal one at the same temperature and density can be written approximately as the virial equation: $P_{\text{real}}-P_{\text{ideal}}=\rho k T \Q(B_2(T)\rho+B_3(T)\rho^2+...\W)$ \cite{huang}. For ideal gases, and in general non-anomalous scale-invariant systems, $2\ma E-DP=0$. Therefore at constant pressure and volume, one can define $\Q(2\ma E_{\text{real}}-DP\W)=\Q(2\ma E_{\text{real}}-DP\W)-\Q(2\ma E_{\text{ideal}}-DP\W)=\Q(2\ma E_{\text{real}}-2\ma E_{\text{ideal}}\W) \equiv 2 \ma E_{\text{res}}$, so that $\Q(2\ma E_{\text{real}}-DP\W)$ equals twice the residual internal energy characterizing the departure of the system from ideal \cite{italians}. In other words, for any system, $2\ma E-DP$ is equal to the difference in its energy from any non-anomalous scale-invariant system's energy at the same $V$ and $P$. For ultracold gases interacting via contact interaction, $2\ma E-DP$ is proportional to the Tan contact $ \lambda^2\la \psi^\da_\uparrow\psi^\da_\downarrow \psi_\downarrow \psi_\uparrow \ra$ \cite{tanpressure}. Many universal relations depending only on the contact exist, independent of the exact details of the experimental setup \cite{tannal,tanrelation}. \\

For systems that are scale invariant at the level of the classical action, a non-zero value of $2\ma E-DP$ signifies a quantum anomaly, so that $2\ma E-DP$ measures quantum anomalies. Previously, it was shown that even in anomalous non-relativistic systems, $2\ma E-DP$ can be expressed as a functional determinant via use of Fujikawa's path integral methods \cite{ord,nonamer,lin,46}. Therefore, one can potentially extract information about $\beta(C)$ and hence obtain information from or even solve the scattering problem by extracting information from the thermodynamic problem.   \\

In this paper we derive a simple expression for $2\ma E-DP$ from dimensional analysis and thermodynamics, independent of quantum mechanics or field theory and independent of Noether's theorem and canonical commutation relations. In this paper we will use units where $\hbar=m=k_B=1$. The units for all quantities can then be written as $\hbar^i m^j k_B^k L^\ell=L^\ell$, where $L$ is a variable in the problem with units of length. We will define $[g_k]=\ell$, and call $\ell$ the dimensions of the variable $g_k$. \\

With this convention, $2\ma E-DP=\sum\limits_{k}\,[g_k]g_k\pf{P}{g_k}$, where $\ma E=\f{E}{V}$ is the thermal energy per unit volume, $P$ is the pressure, and $D$ is the number of spatial dimensions. In this formula $g_k$ are the microscopic parameters of the theory, and $[g_k]$ are the dimensions of these parameters. The derivatives w.r.t. microscopic parameters are taken at constant temperature $\beta^{-1}$, volume $V$, and chemical potential for each species $\mu_i$. The LHS is written in terms of pure macroscopic thermodynamic variables, while the RHS contains derivatives purely on the microscopic parameters. Such an equation can be seen as connecting thermodynamics on the LHS (variables characterizing the macrostate) and statistical mechanics on the RHS (microscopic variables that are system dependent). In particular, for a theory in which all the couplings are dimensionless (in the sense that they have no length dimension as defined above), $[g_k]=0$, and one might expect the system to be scale invariant with $2\ma E-DP=0$. However, for such systems, we show  $2\ma E-DP=-\beta(C)\pf{P}{C}$. The microscopic parameters $g_k$ of a system usually appear in its Hamiltonian as coupling constants, except in the case of dimensional transmutation. The latter leads to a new scale appearing in the pressure $P$, and in the literature this is called a quantum anomaly \cite{norway}. Therefore $2\ma E-DP$ is also a measure of the anomaly for classically scale-invariant systems.\\

The relativistic generalization is $\ma E-DP=\sum\limits_{k}\,[g_k]g_k\pf{P}{g_k}$. It was shown in \cite{callan,callans} that the trace of the improved stress-energy tensor in relativistic $\lambda \phi^4$ has the property $\theta^{00}-\sum\limits_{i}\theta^{ii}=m^2\phi^2$, where the mass term represents a dilational symmetry-breaking term and the improved stress-energy tensor $\theta^{\mu \nu}$ is related to the canonical one $T^{\mu \nu}$ by $\theta^{\mu \nu}=T^{\mu \nu}+\f{D-1}{4D}\Q(g^{\mu \nu}\partial^2-\partial^\mu\partial^\nu\W)\phi^2$ \cite{japan}. Identifying $\theta^{00}$ as $\ma E$ and $\sum\limits_{i}\theta^{ii}=D P_H$, where $P_H$ is the hydrodynamic pressure, one derives the thermal analog. $\theta^{ii}$ is equal to the hydrodynamic pressure \cite{pressure}: however, in equilibrium, the thermodynamic pressure $P$ equals $P_H $ via the virial theorem (although anomalies can complicate matters \cite{weert}). Therefore deriving this expression from the stress-energy tensor requires an improvement of the stress-energy tensor, and an identification of field variables with thermodynamic variables. Similarly, the canonical stress-tensor $T_{ij}$ for nonrelativistic fields can be improved to $\theta_{ij}=T_{ij}+\f{D}{4(D-1)}\Q(\delta_{ij}\nabla^2-\partial_i\partial_j\W)(\psi^\da \psi)$ such that $2\ma \theta_{00}-\theta_{ii}$, which can be identified with $2\ma E-DP$, gives the divergence of the dilation current \cite{hag}. We avoid the complications of having to construct the improved stress-energy tensor, or having to work in the context of field theory, by working directly within thermodynamics. We show the consistency of the equation for a variety of cases, with and without anomalies, and then we show that starting from $2\ma E-DP=\sum\limits_{k}\,[g_k]g_k\pf{P}{g_k}$, one can derive the virial theorem, further illustrating the correctness of the expression and showing the relationship between scaling and the virial theorem. The relativistic case is also considered. \\

While the emphasis of this paper is on structural aspects of the thermodynamical traces $2\ma E-DP$ and $\ma E-DP$ (nonrelativistic and relativistic cases respectively), we have included in section \ref{sec4} a few examples that connect our results with the literature and illustrate how to apply the techniques developed here. In our recent work \cite{ord,virpap,nonamer,lin}  we have aimed at offering a different approach and new perspectives to  the study of the  type of systems presented here. We hope to use the insight gained to apply these techniques and concepts to other problems and systems \cite{prog}.

\section{Proof of Relation} 

For ease of presentation, take the independent, dimensionful microscopic parameters $g_k$ of our theory, and form new parameters $E_k$ with dimensions of energy by raising $g_k$ to the appropriate power, and rewrite the pressure in terms of these new variables\footnote{e.g. if one has a scattering length $a$, replace it with the variable $E_k=1/a^2=\hbar/ma^2$.}. The grand potential $\Omega=\Omega(\beta, \mu_i, V, E_i)$ for a homogeneous system in $D$-spatial dimensions must have the form

\en \label{1}
\Omega(\beta, z_i, V, E_i)=V \beta^{-1-\f{D}{2}} f(z_i,\beta E_i),
\een

where $f(z_i,\beta E_i)$ is a dimensionless function of dimensionless variables, and $z_i=e^{\beta \mu_i}$ is the fugacity corresponding to $\mu_i$. The reason $\Omega$ must have this form is because $\beta$ and $\mu_i$ don't depend on the absolute size of the system (they are intensive variables). If one doubles the system keeping $\beta$ and $\mu_i$ constant, then $\Omega$, being an extensive quantity, should double. So $\Omega$ must be proportional to V. To make up for the remaining dimension ($[\Omega]=-2$), we are free to pull out one of the dimensionful arguments of $\Omega$, and the rest of the arguments must be ratios with the argument we pulled out. We will pull out $\beta$. This is equivalent to choosing our scale as $\beta$ and measuring all other quantities in units of $\beta$.\\

Now take the derivative of Eq. \eqref{1} w.r.t. to $\beta$ at constant fugacity $z_i$ and volume $V$, and multiply times $\beta$:

\en \label{2}
\begin{split}
\left.\beta \pf{\Omega}{\beta}\right|_{z_i,V}&=\Q(-1-\f{D}{2}\W)\Omega+V \beta^{-1-\f{D}{2}} \beta \left.\pf{ f(z_i,\beta E_i)}{\beta}\right|_{z_i}\\
&=\Q(-1-\f{D}{2}\W)\Omega+V \beta^{-1-\f{D}{2}} \beta \Q[\sum\limits_{k}\f{E_k}{\beta} \left.\pf{ f(z_i,\beta E_i)}{E_k}\W]\right|_{z_i}\\
&=\Q(-1-\f{D}{2}\W)\Omega+\sum\limits_{k}E_k \pf{\Omega}{E_k}.
\end{split}
\een

Now, we use the thermodynamic identity $E=\left. \pf{(\beta \Omega)}{\beta}\right|_{z_i,V}=\Omega+\left .\beta \pf{ \Omega}{\beta}\right|_{z_i,V}$. 
\en \label{3}
\begin{split}
2E-DPV&=2\Q(\Omega+\left .\beta \pf{ \Omega}{\beta}\right|_{z_i,V} \W)-DPV\\
&=2\Q(\Omega+\Q(-1-\f{D}{2}\W)\Omega+\sum\limits_{k} E_k \pf{\Omega}{E_k} \W)-DPV\\
&=-2\Q(P+\Q(-1-\f{D}{2}\W)P+\sum\limits_{k} E_k \pf{P}{E_k} \W)V-DPV\\
&=-2\sum\limits_{k}E_k \pf{P}{E_k}V \\
2\ma E-DP&=-2\sum\limits_{k} E_k \pf{P}{E_k}. 
\end{split}
\een

For 0-temperature, we lose $\beta$ as a scale. Instead we use $\mu_1$, where $\mu_1$ is the chemical potential for one of the particles:

\en \label{4}
\Omega=V\mu_1^{1+D/2}f\left(\f{\mu_1}{E_i},\f{\mu_1}{\mu_{j\neq 1}}\right),
\een

The calculation is done in appendix \ref{app2}.\\
 
In general, as long as the theory has microscopic parameters $g_i$ that have dimensions of length (and not necessarily energy or $L^{-2}$, then by forming appropriate dimensionless variables $x_i=\beta^{-\f{[g_i]}{2}} g_i$ for the argument of $\Omega(\beta, z, V, g_i)=V \beta^{-1-\f{D}{2}} f(z,\beta^{-\f{[g_i]}{2}}g_i)$, then one gets:

\en \label{7}
2\ma E-DP&=\sum\limits_{k}[g_k]g_k\pf{P}{g_k}.
\een

Alternatively, one can note that $E_k=g_k^{-\f{2}{[g_k]}}$, and apply the chain rule to Eq. \eqref{3} to get Eq. \eqref{7}.

\section{Relativistic Systems}

In relativistic theories, $\hbar=c=k_B=1$, and mass attains a dimension equal to $1/L$. 
The units for all quantities can then be written as $\hbar^ic^jk_B^kL^\ell=L^\ell$, and we define the dimensions of the parameter $g_k$ as $[g_k]=\ell$. The grand potential $\Omega$ has $[\Omega]=-1$ rather than the NR case $[\Omega]=-2$, and can be written as:

\en \label{8a}
\Omega(\beta, z_i, V, E_i)=V \beta^{-1-D} f(z_i,\beta E_i).
\een

Following more or less the same steps as before one derives:

\en \label{9a}
\ma E-DP&=\sum\limits_{k}\,[g_k]g_k\pf{P}{g_k},
\een

where again as in the nonrelativistic case, the derivatives are taken w.r.t. constant $\beta^{-1}$, $V$, and $\mu_i$.

\section{Examples}
\label{sec4}
\subsection{No anomalies, no dimensionful parameters}

The Tonks-Girardeau gas \cite{tonks,gord} is a 1-dimensional gas of bosons interacting via a two-body contact potential $V=2g\delta(x_i-x_j)$ in the limit $g \rightarrow \infty$. At zero temperature, $\ma E=\f{\pi^2}{6}\rho^3$ and $P=\f{\pi^2}{3}\rho^3$. Since the only dimensional parameter grows without bound, the gas is scale invariant, and hence by Eq. \eqref{7}:

\en  \label{8}
2\ma E-P=0.
\een 

\subsection{No anomalies, dimensionful parameters}

For a contact-interaction Bose gas at zero temperature (i.e. $\ma L= \psi^\da \Q(i\p_t+\f{\nabla^2}{2}\W)\psi-\f{g}{2} \Q(\psi^\da \psi \W)^2)$, in odd dimensions $D=2n+1$ (perfectly finite in dimensional regularization, no anomalies), one can make the following 1-loop calculation for small coupling \cite{bbs,bratnet}

\en \label{9}
\Omega=\Q(-\f{1}{2}\f{\mu^2}{g}-L_D \, \mu^{\f{D}{2}+1}\W)V,
\een

where $\Omega$ is the grand potential, $L_D=\f{\Gamma\Q(1-\f{D}{2}\W)\Gamma\Q(\f{D+1}{2}\W)}{2\pi^{\f{D+1}{2}}\Gamma\Q(\f{D}{2}+2\W)}$ is a pure number that depends only on dimension. We will verify Eq. \eqref{7} by computing the LHS involving macroscopic thermodynamic parameters by using thermodynamic relations on Eq. \eqref{9}. Then we will calculate the LHS of Eq. \eqref{7} by differentiation w.r.t. microscopic parameters of Eq. \eqref{9}, and compare the two results. \\

For the LHS, the following thermodynamic identities will be used, true for any homogeneous system: 

\en \label{10}
\begin{split}
\Omega=-PV, \\
 \Omega=E-TS-\mu N \Rightarrow E=\Omega+\mu N, \\ 
N=-\pf{\Omega}{\mu}.
\end{split}
\een

Calculating $N$ for Eq. \eqref{9} using Eq. \eqref{10}:

\en \label{11}
N=\Q(\f{\mu}{g}+L_D \,\Q(\f{D}{2}+1\W) \mu^{\f{D}{2}}\W)V.
\een

Therefore:

\en \label{12}
\begin{split}
2E-DPV&=2(\mu N-PV)-DPV=2\mu N-(D+2)PV\\
&=2\mu \Q(\f{\mu}{g}+L_D \,\Q(\f{D}{2}+1\W) \mu^{\f{D}{2}}\W)V+(D+2)\Q(-\f{1}{2}\f{\mu^2}{g}-L_D \, \mu^{\f{D}{2}+1}\W)V \\
&=\Q(\Q[1-\f{D}{2}\W] \f{\mu^2}{g}\W)V\\
2\ma E-DP&=\Q(\Q[1-\f{D}{2}\W] \f{\mu^2}{g}\W).
\end{split}
\een

Now make the same calculation but using the microscopic scales. Since we restrict ourselves to $D=2n+1$, there is no renormalization scale as everything is perfectly finite, a feature peculiar to odd dimensions \cite{pseudo}. However, there is a microscopic length scale associated with the coupling $g$, where $[g]=D-2$:

\en \label{13}
\pf{P}{g}[g]g=\f{-\p\Q(\f{\Omega}{ V}\W)}{\p g}(D-2)g=\Q(\Q[1-\f{D}{2}\W] \f{\mu^2}{g}\W).
\een

The above calculation was performed for small coupling $g$ so that perturbation theory could be used, but the relationship is in fact general. The Lieb-Liniger model which describes N bosons in one dimension interacting via a two-body contact potential interaction $H=-\sum\limits_{i}^{N}\f{1}{2}\f{\p^2}{\p x_i^2}+ \sum \limits_{i<j} 2g\delta(x_i-x_j)$ is exactly solveable in quantum mechanics, and the thermodynamic limit $N,L \rightarrow \infty$ with $\f{N}{L}=\rho=\text{constant}$ can be taken \cite{lieb}. Approximate closed-form solutions exist for large and small $g$ at zero temperature. For large coupling:

\en  \label{aa1}
\ma E&=\frac{\sqrt{2}\mu ^{3/2}}{3
   \pi }+\frac{4 \mu ^2}{3 \pi ^2 g}+\frac{14 \mu ^{5/2}}{3 \sqrt{2}\pi ^3 g^2}+\frac{32 \left(100-9 \pi ^2\right) \mu ^3}{405 \pi ^4 g^3}, \\
P&=\frac{2\sqrt{2} \mu ^{3/2}}{3 \pi }+\frac{4 \mu ^2}{3 \pi ^2 g}+\frac{14 \sqrt{2} \mu ^{5/2}}{9 \pi ^3 g^2}+\frac{16 \left(100-9 \pi ^2\right) \mu ^3}{405 \pi ^4 g^3},\\
-g\f{\p P}{\p g}&=\frac{4 \mu ^2}{3 \pi ^2 g^2}+\frac{28 \sqrt{2} \mu ^{5/2}}{9 \pi ^3 g^3}+\frac{16 \left(100-9 \pi ^2\right) \mu ^3}{135 \pi ^4 g^4}.
\een

In the limit $\f{g}{\sqrt{\mu}} \rightarrow \infty$, the residual energy goes to zero and one gets the Tonks-Girardeau gas of Eq. \eqref{8}.\\

For small coupling:

\en \label{aa2}
\ma E&=\frac{\mu ^2}{4 g}+\frac{\mu ^{3/2}}{3 \pi },\\
P&=\frac{\mu ^2}{4 g}+\frac{2 \mu ^{3/2}}{3 \pi },\\
-g\f{\p P}{\p g}&=\frac{\mu ^2}{4 g},
\een

which agrees with the Bogoliubov approximation of Eq. \eqref{12} when $g\rightarrow\f{g}{2}$ and $D=1$.\\

For fermions in 3-dimensions interacting via contact interactions $\ma L= \psi^\da \Q(i\p_t+\f{\nabla^2}{2}\W)\psi-4\pi a\psi^\da_\uparrow\psi^\da_\downarrow \psi_\downarrow \psi_\uparrow$, $[a]=1$:

\en  \label{14a}
2\ma E-3P=[a]a\f{\p P}{\p a}.
\een

Now $\beta PV=\ln \int [d\psi d\psi^\da] e^{-\int^\beta_0 \int_V d\tau d^2x\, \Q(\ma L_0+4\pi a\psi^\da_\uparrow\psi^\da_\downarrow \psi_\downarrow \psi_\uparrow\W)}$ so that differentiating the path integral w.r.t. $a$ gives

\en 
\,[a]a\f{\p P}{\p a}=-4\pi a\la \psi^\da_\uparrow\psi^\da_\downarrow \psi_\downarrow \psi_\uparrow \ra.
\een

Plugging this into Eq. \eqref{14a}, we get Tan's pressure relation:

\en \label{qwerty}
2\ma E-3P=-4\pi a\la \psi^\da_\uparrow\psi^\da_\downarrow \psi_\downarrow \psi_\uparrow \ra=-\f{C}{4\pi a},
\een

where $C=(4\pi a)^2\la \psi^\da_\uparrow\psi^\da_\downarrow \psi_\downarrow \psi_\uparrow \ra $ is the Tan contact \cite{tanpressure}. Eq. \eqref{qwerty} was also derived using the Hellmann-Feynman theorem and dimensional arguments in the canonical ensemble in \cite{braa}. Indeed, the energy $E$, Helmholtz energy $F$, Gibbs energy $G$, and grand potential $\Omega$ are related by Legendre transformation that trades conjugate macroscopic variables but leaves the microscopic parameters alone. Therefore:

\en
\left.\f{\p \Omega}{\p g}\right|_{\mu,V,T}=\left.\f{\p F}{\p g}\right|_{N,V,T}=\left.\f{\p G}{\p g}\right|_{N,P,T}=\left.\f{\p E}{\p g}\right|_{S,V,T}.
\een

 In fact, for the Lieb-Liniger model, where the N-body system is exactly solveable, it is more natural to work with the density $\rho$ instead of the chemical potential $\mu$:

\en
\ma E&=\frac{1}{6} \pi ^2 \left(1-\frac{4}{\gamma }+\frac{12}{\gamma ^2}+ \frac{32 \left(\pi ^2-15\right)}{15 \gamma ^3}\right) \rho ^3 \quad \text{(strong)},\\
\ma E&=\f{1}{2}\left(\gamma-\frac{4 \gamma^{3/2}}{3 \pi }\right)   \rho ^3  \quad \text{(weak)},
\een 

where $\gamma=\f{2g}{\rho}$. By the third law of thermodynamics, along a zero temperature path, variations in the coupling $\gamma$ occur at constant entropy $S$, so twice the residual energy is:

\en
[g]g\f{\p E}{\p g}=\gamma\f{\p E}{\p \gamma}&=\frac{1}{6} \pi ^2 \left(\frac{4}{\gamma }-\frac{24}{\gamma^2}- \frac{32 \left(\pi ^2-15\right)}{5 \gamma^3}\right) \rho ^3=g\Q \la \Q(\psi^\da (x)\W)^2\Q(\psi(x)\W)^2\W \ra   \quad \text{(strong)},\\
&=\f{1}{2}\left(\gamma-\frac{2 \gamma^{3/2}}{ \pi }\right)   \rho ^3 =g \Q \la \Q(\psi^\da (x)\W)^2\Q(\psi(x)\W)^2\W \ra  \quad \text{(weak)}.
\een

\subsection{Anomalies, no dimensionful parameters}

A Fermi-gas in $D=2$ has no dimensionful parameters in the Lagrangian, $\ma L= \psi^\da \Q(i\p_t+\f{\nabla^2}{2}\W)\psi-C\psi^\da_\uparrow\psi^\da_\downarrow \psi_\downarrow \psi_\uparrow$, $[C]=0$. Nevertheless, the system develops a bound state via dimensional transmutation, and the derivation of Eq. \eqref{3} goes through with $E_k=E_b$, the bound state energy. Using cutoff regulariztion, the $T$-matrix is \cite{tmatrix}:
\en \label{14}
\f{1}{T(E)}=\f{1}{C}-\f{1}{4\pi} \ln\Q(\f{-E}{\Lambda^2}\W).
\een

The bound state is special since $T(E)$ blows up there, so that $\f{1}{T(-E_b)}=0$. Therefore plugging in $E=-E_b$ into Eq. \eqref{14} gives:

\en \label{15}
\f{1}{C}=\f{1}{4\pi} \ln\Q(\f{E_b}{\Lambda^2}\W).
\een

Taking the derivative w.r.t. $E_b$ on both sides of Eq. \eqref{15}:

\en \label{16}
-\f{\f{dC}{dE_b}}{C^2}=\f{1}{4\pi}\f{1}{E_b}\\
\f{dC}{dE_b}=-\f{C^2}{4\pi}\f{1}{E_b}.
\een

Therefore:

\en \label{17}
-2E_b \f{\p P}{\p E_b}=-2E_b \f{dC}{dE_b} \f{\p P}{\p C}=\f{C^2}{2\pi}\f{\p P}{\p C}.
\een

Now $\beta PV=\ln \int [d\psi d\psi^\da] e^{-\int^\beta_0 \int_V d\tau d^2x\, \Q(\ma L_0+C\psi^\da_\uparrow\psi^\da_\downarrow \psi_\downarrow \psi_\uparrow\W)}$ so that differentiating the path integral w.r.t. $C$ we obtain:

\en \label{18}
\f{\p P}{\p C}=-\la \psi^\da_\uparrow\psi^\da_\downarrow \psi_\downarrow \psi_\uparrow \ra.
\een

Plugging this result into Eq. \eqref{17} and using $2\ma E-DP=-2E_b \pf{P}{E_b}$:

\en \label{19}
2\ma E-2P=-\f{C^2}{2\pi}\la \psi^\da_\uparrow\psi^\da_\downarrow \psi_\downarrow \psi_\uparrow \ra,
\een

agreeing with \cite{hoff}. The coupling is bare, but the RHS is finite, and both sides are RG-invariant.\\

In our example, for Eq. \eqref{7}, the dimensionful parameter is the bound-state energy. If one has a pressure written in term of bare parameters and cutoff $P=P(C,\Lambda)$ or renormalized with scale $\mu$, $P=P(C_R, \mu)$, then it is not correct to regard $\Lambda$ or $\mu$ as a microscopic parameter with dimensions of momentum ($L^{-1}$), because $\f{dP}{d\Lambda}=\f{dP}{d\mu}=0$, so that there is in fact no dependence on these parameters. For our particular example, from Eq. \eqref{15}, it is true that $2E_b \f{dC}{dE_b}=-\Lambda \f{dC}{d\Lambda}=-\beta(C)$ where $\beta(C)$ is the beta function of the theory, so that Eq. \eqref{17} into our Eq. \eqref{7} would give:

\en \label{20a}
2\ma E-DP=- \beta(C) \f{\p P}{\p C},
\een 

and comparison with Eq. \eqref{19} allows us to read off $\beta(C)=\f{C^2}{2\pi}$. \\

\section{Connection with Virial Theorem}

In previous work \cite{virpap} we derived the virial theorem via path integrals, and then used the virial theorem to derive Eq. \eqref{7}. One can also work backwards from Eq. \eqref{7} to derive the virial theorem by following the argument backwards. We reproduce the argument here. For a two-body potential  $U=\f{1}{2} \int d^Dxd^Dy \psi^*(\tau,\vec{x}) \psi(\tau,\vec{x})  V(\vec{x}-\vec{y}\,)   \psi^*(\tau,\vec{y}) \psi(\tau,\vec{y})$:

\en \label{20}
2&E-DPV=V\sum\limits_{k}\,[g_k]g_k\pf{P}{g_k}\\
&=\sum\limits_{k}\, [g_k]g_k\f{1}{\beta}\p_{g_k}\ln \int [d\psi d\psi^\da] e^{-\int^\beta_0 \int_V d\tau d^Dx\, \Q(\ma L_0+\f{1}{2} \int d^D \vec{y} \,\psi^*(\tau,\vec{x}) \psi(\tau,\vec{x})V(\vec{x}-\vec{y}\,)\psi^*(\tau,\vec{y}) \psi(\tau,\vec{y})\W)}\\
&=\sum\limits_{k}\,[g_k]g_k \Q(\f{-1}{2}\W)  \Q \la \int_V\int_V d\tau d^Dx d^Dy  \psi^*(\tau,\vec{x}) \psi(\tau,\vec{x}) \f{\p V}{\p g_k}  \psi^*(\tau,\vec{y}) \psi(\tau,\vec{y}) \W \ra.
\een

Denoting $r=|\vec{x}-\vec{y}\,|$, one can show that $-\sum\limits_{k}\,[g_k]g_k\f{\p V}{\p g_k}=r\f{dV}{dr}+2V$ (see appendix \ref{app1}). Plugging this into Eq. \eqref{20} gives:

\en \label{21}
2E-DPV&=\f{1}{2} \Q \la  \psi^*(\tau,\vec{x}) \psi(\tau,\vec{x}) r\f{dV}{dr}  \psi^*(\tau,\vec{y}) \psi(\tau,\vec{y}) \W \ra +2\Q \la U \W\ra \\
DPV&=2KE- \\ &\f{1}{2}\Q\la \int d^Dxd^Dy \, \psi^*(\tau,\vec{x}) \psi(\tau,\vec{x}) \Q[(\vec{x}-\vec{y}\,)\cdot \na_{\vec{x}} V(\vec{x}-\vec{y}\,)  \W]  \psi^*(\tau,\vec{y}) \psi(\tau,\vec{y}) \W \ra,
\een
 
which is the virial theorem \cite{t2,59,61}. \\

\section{Conclusion} 

We have derived an expression for $2\ma E-DP$ using only dimensional arguments, valid for classical and quantum systems, for use in the grand canonical ensemble. We worked directly within the framework of thermodynamics, not having to improve the stress-energy tensor and invoke hydrodynamics, but instead working directly with thermodynamic variables. In the case of quantum systems, since the microscopic scales appear as coupling constants, or in the case of dimensional transmutation appear via the coupling constants, $\sum\limits_{k}\, [g_k]g_k\pf{P}{g_k}$manifests itself as thermal expectation values of the operators multiplying the coupling constants in the system's Hamiltonian, which is manifest in the path integral formalism. Finally, using the path integral, we've shown how dimensional analysis leads to the virial theorem.       \\

\begin{center} \textbf{\small Conflict of Interests} \end{center} 

The authors declare that there is no conflict of interests regarding the publication of this paper.

\begin{acknowledgements}

The authors would like to thank the reviewer for many helpful suggestions. This work was supported in part by the US Army Research Office Grant No. W911NF-15-1-0445.

\end{acknowledgements}

\appendix

\section{}
\label{app1}
The potential $V(r)$ has dimensions $[V]=-2$, so can generically be written:

\en
V(r)=\f{f\left(\f{g_i}{r^{[g_i]}}\right)}{r^2}.
\een

$f$ is a dimensionless function whose arguments are the ratios of the couplings $g_i$ of $V(r)$ to their length dimension $[g_i]$ expressed in units of $r$.

\en
r\f{dV}{dr}&=-2V(r)+\f{1}{r}\f{df\left(\f{g_i}{r^{[g_i]}}\right)}{dr} \\
&=-2V(r)-\f{1}{r^2} \sum\limits_{i}\, [g_i]g_i\f{\p  f\left(\f{g_i}{r^{[g_i]}}\right)}{\p g_i}\\
&=-2V(r)-\sum\limits_{i}\,[g_i] g_i \f{\p V}{\p g_i}.
\een

\section{} 
\label{app2}
At zero temperature for a homogeneous system, the grand potential $\Omega$ can be written as:

\en 
\Omega=V\mu_1^{1+D/2}f\left(\f{\mu_1}{E_i},\f{\mu_1}{\mu_{j\neq 1}}\right),
\een

where $f$ is dimensionless function and $\mu_1$ is the non-zero chemical potential of one of the species. We calculate the number of particles:

\en \label{5}
\begin{split}
N_1&=-\pf{\Omega}{\mu_1}\Big |_{V, \mu_{j\neq 1}}\\
&=-(1+D/2) \f{\Omega}{\mu_1}-V\mu_1^{1+D/2}\pf{f\left(\f{\mu_1}{E_j},\f{\mu_1}{\mu_{j\neq 1}} \right)}{\mu_1}\\
&=-(1+D/2) \f{\Omega}{\mu_1}\\&-V\mu_1^{1+D/2}\left[-\sum\limits_{k}\f{E_k}{\mu_1}\pf{f\left(\f{\mu_1}{E_j},\f{\mu_1}{\mu_{j\neq 1}}\right)}{E_k}-\sum\limits_{\ell\neq1}\f{\mu_{\ell\neq 1}}{\mu_1}\pf{f\left(\f{\mu_1}{E_j},\f{\mu_1}{\mu_{j\neq 1}}\right)}{\mu_{\ell\neq 1}}\right]\\
&=-(1+D/2) \f{\Omega}{\mu_1}+\sum\limits_{k}\f{E_k}{\mu_1}\pf{}{E_k}\Omega+\sum\limits_{\ell \neq 1}\f{\mu_{\ell\neq 1}}{\mu_1}\pf{}{\mu_{\ell\neq 1}}\Omega,\\
N_1\mu_1&=-(1+D/2)\Omega+\sum\limits_{k}E_k\pf{}{E_k}\Omega-\sum\limits_{\ell \neq 1}N_{\ell\neq 1}\mu_{\ell\neq 1},\\
\sum\limits_{i}N_i\mu_i&=-(1+D/2)\Omega+\sum\limits_{k}E_k\pf{}{E_k}\Omega.
\end{split}
\een

The energy $E$ of the system at zero temperature is given by $E=\sum\limits_{i} N_i \mu_i-PV$ which follows from the thermodynamic identity $E-TS+PV=\sum\limits_{i} N_i \mu_i$. Therefore

\en \label{6}
\begin{split}
2E-DPV&=2\Q(\sum\limits_{i}N_i \mu_i-PV\W)-DPV=2\sum\limits_{i} N_i \mu_i-(D+2)PV\\
&=2\left( -(1+D/2)\Omega+\sum\limits_{k}E_k\pf{}{E_k}\Omega \right)-(D+2)PV \\
&=2\left( -(1+D/2)(-PV)+\sum\limits_{k}E_k\pf{}{E_k}(-PV) \right)-(D+2)PV\\
&=-2V\sum\limits_{k}E_k\pf{}{E_k}P,\\
2\ma E-DP&=-2\sum\limits_{k}E_k\pf{P}{E_k}.
 \end{split}
\een


\begin{thebibliography}{10}

\bibitem{brown}
L.~Brown, {\em {Quantum field theory}}.
\newblock Cambridge Univ. Pr., 1992.

\bibitem{car}
P.~Carruthers, ``Broken scale invariance in particle physics,'' {\em Physics
  Reports}, vol.~1, no.~1, pp.~1 -- 29, 1971.

\bibitem{gat}
J.~Gaite, ``The relativistic virial theorem and scale invariance,'' {\em
  Physics-Uspekhi}, vol.~56, no.~9, p.~919, 2013.

\bibitem{huang}
K.~Huang, {\em {Statistical Mechanics}}.
\newblock Wiley, 1987.

\bibitem{italians}
F.~Mancarella, G.~Mussardo, and A.~Trombettoni, ``{Energy-pressure relation for
  low-dimensional gases},'' {\em Nucl.Phys.}, vol.~B887, pp.~216--245, 2014.

\bibitem{tanpressure}
S.~Tan, ``Generalized virial theorem and pressure relation for a strongly
  correlated fermi gas,'' {\em Annals of Physics}, vol.~323, no.~12, pp.~2987
  -- 2990, 2008.

\bibitem{tannal}
S.~Tan, ``Energetics of a strongly correlated fermi gas,'' {\em Annals of
  Physics}, vol.~323, no.~12, pp.~2952 -- 2970, 2008.

\bibitem{tanrelation}
S.~Tan, ``Large momentum part of a strongly correlated fermi gas,'' {\em Annals
  of Physics}, vol.~323, no.~12, pp.~2971 -- 2986, 2008.

\bibitem{ord}
{C.R. Ord\'o\~nez}, ``Path-integral fujikawa’s approach to anomalous virial
  theorems and equations of state for systems with symmetry,'' {\em Physica A:
  Statistical Mechanics and its Applications}, vol.~446, pp.~64 -- 74, 2016.

\bibitem{nonamer}
C.~L. Lin and C.~R. Ord\'o\~nez, ``Bose and fermi statistics and the
  regularization of the nonrelativistic jacobian for the scale anomaly,'' {\em
  Phys. Rev. D}, vol.~94, p.~085001, Oct 2016.

\bibitem{lin}
C.~L. Lin and C.~R. Ord\'o\~nez, ``Path-integral derivation of the
  nonrelativistic scale anomaly,'' {\em Phys. Rev. D}, vol.~91, p.~085023, Apr
  2015.

\bibitem{46}
C.~L. Lin and C.~R. Ord\'o\~nez, ``Path-integral approach to scale anomaly at
  finite temperature,'' {\em Phys. Rev. D}, vol.~92, p.~085050, Oct 2015.

\bibitem{norway}
T.~Haugset and F.~Ravndal, ``{Scale anomalies in nonrelativistic field theories
  in (2+1)-dimensions},'' {\em Phys.Rev.}, vol.~D49, pp.~4299--4301, 1994.

\bibitem{callan}
S.~Coleman and R.~Jackiw, ``Why dilatation generators do not generate
  dilatations,'' {\em Annals of Physics}, vol.~67, no.~2, pp.~552 -- 598, 1971.

\bibitem{callans}
C.~G. Callan, S.~Coleman, and R.~Jackiw, ``A new improved energy-momentum
  tensor,'' {\em Annals of Physics}, vol.~59, no.~1, pp.~42 -- 73, 1970.

\bibitem{japan}
K.~Shizuya and H.~Tsukahara, ``Path-integral formulation of conformal
  anomalies,'' {\em Zeitschrift f{\"u}r Physik C Particles and Fields},
  vol.~31, no.~4, pp.~553--556, 1986.

\bibitem{pressure}
P.~Jizba, ``Hydrostatic pressure of the $o(n)$ ${\ensuremath{\varphi}}^{4}$
  theory in the large \textit{N} limit,'' {\em Phys. Rev. D}, vol.~69,
  p.~085011, Apr 2004.

\bibitem{weert}
N.~P. {Landsman} and C.~G. {van Weert}, ``{Real- and imaginary-time field
  theory at finite temperature and density},'' {\em physrep}, vol.~145,
  pp.~141--249, Jan. 1987.

\bibitem{hag}
C.~R. Hagen, ``Scale and conformal transformations in galilean-covariant field
  theory,'' {\em Phys. Rev. D}, vol.~5, pp.~377--388, Jan 1972.

\bibitem{virpap}
C.~L. Lin and C.~R. Ord\'o\~nez, ``{Virial Theorem for Non-relativistic Quantum
  Fields in D Spatial Dimensions},'' {\em Adv. High Energy Phys.}, vol.~2015,
  p.~796275, 2015.

\bibitem{prog}
{Daza W. S., Lin C. L. and Ord\'o\~nez C.R.}, ``{In progress}.''

\bibitem{tonks}
L.~Tonks, ``The complete equation of state of one, two and three-dimensional
  gases of hard elastic spheres,'' {\em Phys. Rev.}, vol.~50, pp.~955--963, Nov
  1936.

\bibitem{gord}
M.~Girardeau, ``Relationship between systems of impenetrable bosons and
  fermions in one dimension,'' {\em Journal of Mathematical Physics}, vol.~1,
  no.~6, pp.~516--523, 1960.

\bibitem{bbs}
A.~Schakel, {\em Boulevard of Broken Symmetries: Effective Field Theories of
  Condensed Matter}.
\newblock World Scientific, 2008.

\bibitem{bratnet}
E.~Braaten and A.~Nieto, ``Quantum corrections to the energy density of a
  homogeneous bose gas,'' {\em The European Physical Journal B - Condensed
  Matter and Complex Systems}, vol.~11, no.~1, pp.~143--159, 1999.

\bibitem{pseudo}
K.~W\'odkiewicz, ``Fermi pseudopotential in arbitrary dimensions,'' {\em Phys.
  Rev. A}, vol.~43, pp.~68--76, Jan 1991.

\bibitem{lieb}
E.~H. Lieb and W.~Liniger, ``Exact analysis of an interacting bose gas. i. the
  general solution and the ground state,'' {\em Phys. Rev.}, vol.~130,
  pp.~1605--1616, May 1963.

\bibitem{braa}
E.~Braaten, {\em Universal Relations for Fermions with Large Scattering Length,
  edited by Zwerger W.}
\newblock Berlin, Heidelberg: Springer Berlin Heidelberg, 2012.

\bibitem{tmatrix}
D.~R. Phillips, S.~R. Beane, and T.~D. Cohen, ``Nonperturbative regularization
  and renormalization: Simple examples from nonrelativistic quantum
  mechanics,'' {\em Annals of Physics}, vol.~263, no.~2, pp.~255 -- 275, 1998.

\bibitem{hoff}
J.~Hofmann, ``Quantum anomaly, universal relations, and breathing mode of a
  two-dimensional fermi gas,'' {\em Phys. Rev. Lett.}, vol.~108, p.~185303, May
  2012.

\bibitem{t2}
T.~Toyoda and K.~ichi Takiuchi, ``Quantum field theoretical reformulation of
  the virial theorem,'' {\em Physica A: Statistical Mechanics and its
  Applications}, vol.~261, no.~3�4, pp.~471 -- 481, 1998.

\bibitem{59}
T.~Toyoda, ``Canonical generator of conformal transformations in
  nonrelativistic quantum many-body systems at finite temperatures,'' {\em
  Phys. Rev. A}, vol.~48, pp.~3492--3498, Nov 1993.

\bibitem{61}
K.~Takiuchi, M.~Okada, H.~Koizumi, K.~Ito, and T.~Toyoda, ``Exact relations for
  two-dimensional electron gas spin correlation functions,'' {\em Physica E:
  Low-dimensional Systems and Nanostructures}, vol.~6, no.~1�4, pp.~810 --
  812, 2000.

\end{thebibliography}
\end{document}